\documentstyle[11pt]{article}
\begin{document}
\begin{titlepage}
\vskip 15 mm
\title{Yukawa textures from an extra {\bf U(1)} symmetry?}
\vskip 10 mm
\author{\bf E. Papageorgiu}
\vskip 2 mm
\centerline{Laboratoire de Physique Th\'{e}orique et Hautes Energies}
\centerline{Universit\'{e} de Paris XI, B\^{a}timent 211, 911405 Orsay, France}
\vskip 20 mm
\abstract{The observed hierarchy of the quark and lepton masses
and mixing angles has led to the parametrisation of the mass
matrices in powers of the Cabibbo angle and a certain number of zeros.
The origin of these ``textures'' may lie in symmetries present
at the GUT/string unification scale. We discuss the case of an extra
$U(1)$. Symmetric textures with two or three zeros can be
generated through high-order operators containing light/heavy
Higgs multiplets and singlet fields. However, a realistic scenario
for generating both the up- and down-quark mass matrices is not possible
within the MSSM with an extra singlet. The presence of at least one pair
of heavy Higgs multiplets and extra singlets are needed,
and, even so, difficulties are
encountered when extending this scenario to the neutrino sector through
the mass relations of grand unification.}

\vskip 1 truecm
\noindent {\bf LPTHE Orsay Preprint 40/94}
\end{titlepage}

The Yukawa sector of the Standard Model (SM)
and its supersymmetric extension (MSSM) contains a large number
of parameters that are related to the fermion masses and the
quark-mixing angles.
Additional assumptions are in particular needed
in order to determine the structure of the $3\times 3$  Yukawa matrices,
which parametrise the couplings of the fermions to the Higgs bosons.
In an attempt to keep the number of input parameters minimal, various
authors have proposed left-right (L-R) symmetric Yukawa matrices
containing two or three zero entries
\footnote{When counting the zeros of a symmetric matrix only the entries
above or below the diagonal are considered.},
so-called ``textures'', which, when multiplied by diagonal matrices of phases,
describe successfully the mass
spectra, and predict relations between the quark masses and the mixing
angles [1-4].
Different {\it Ansatz} are {\it a priori} possible,
not only because of the freedom
of choosing the energy scale at which they are formulated and their
dependence on the particle content of the underlying model,
but also because of present-day experimental uncertainties
\footnote{``Texture zeros'' donot remain zero
throughout the evolution,
but they are referred to as ``zeros'' as long as their actual value
does not affect the masses or mixings to leading order.}.
Lately, the possibility of having L-R symmetric textures that can
correctly parametrise the up- and down-quark mass matrices with a
maximum number of five zeros at
the grand-unification scale $M_G \simeq 10^{16}$ GeV
has received special attention within the
context of the MSSM [2-7]
and a list of
five phenomenologically consistent solutions $(I)$, $(II)$, $(III)$,
$(IV)$ and $(V)$ were presented in
ref.\cite{RRR}. For convenience they have been parametrised in powers of
the Cabibbo angle $\lambda \simeq 0.22$ and the coefficients $a,b,c,d$ and
$a^{\prime},b^{\prime},c^{\prime}$
whose values are given in Table 1:
\begin{equation}
Y_u = \left(
\begin{array}{ccc}
0 & a \lambda^6 & d \lambda^4 \\
a \lambda^6 & b \lambda^4 & c \lambda^2 \\
d \lambda^4 & c \lambda^2 & 1
\end{array}
\right) \,,
\end{equation}
and
\begin{equation}
Y_d = \left(
\begin{array}{rcl}
0 & a^{\prime} \lambda^4 & 0 \\
a^{\prime} \lambda^4 & b^{\prime} \lambda^3 & c^{\prime} \lambda^3 \\
0 & c^{\prime} \lambda^3 & 1
\end{array}
\right) \,.
\end{equation}

While the L-R symmetry appears naturally in some models beyond the SM,
{\it e.g.}, in certain grand-unified (GUT) models that are based on the
$SO(10)$ gauge group, the presence of texture zeros
requires the existence of some extra symmetry, that is not ``family-blind''.
The simplest extension of this type could be an extra $U(1)_X$, under which
all states are charged, but with different quantum numbers to be yet
determined. In this paper we shall investigate which of these textures
could be generated within the framework of such a global or local
symmetry and whether
the fermion charges can be uniquely fixed, imposing eventually
additional constraints, such as the cancellation of
anomalies \cite{Ibanez} and/or the partial unification of Yukawa couplings
\cite{Lang}.

Let us assume that such a $U(1)_X$ symmetry exists in addition to the
$SU(2)_L \times U(1)_Y$ SM symmetries, and that
the quark and lepton
states transform as:
\begin{equation}
q_{i(L,R)} \to e^{i\theta\alpha_i} q_{i(L,R)}\,,
\end{equation}
and
\begin{equation}
l_{i(L,R)} \to e^{i\theta\gamma_i} l_{i(L,R)}\,,
\end{equation}
respectively, where fermions belonging to different families
are supposed to carry different charges, {\it i.e.},
$\alpha_i\not= \alpha_j$ and $\gamma_i\not= \gamma_j$ for $i\not=j$.
Then, if $h^{\chi}$ is the SM Higgs field carrying charge $\chi$,
a second light Higgs doublet is needed to give masses to the down quarks
and charged leptons, unless $\chi = 0$.
However, if $\chi$ is zero, an allowed
$q^c_{i} h q_{j}$ entry in the quark mass matrix would imply opposite
charges for right- and left-handed states. We consider therefore
the two-doublet Higgs case, and denote by $h_1^{\chi}$ the Higgs doublet
that can generate nonzero $(i,j)$
entries in the up-quark mass matrix,
where the following condition for the charges is satisfied:
\begin{equation}
\alpha_i + \alpha_j + \chi = 0\,,
\end{equation}
and by $h_2^{\chi}$ correspondingly the Higgs doublet responsible for zero
and nonzero entries in the down-quark mass matrix.
The above condition implies, that, whenever any two diagonal entries are
allowed, the corresponding off-diagonal entries are allowed too. In addition,
whenever a diagonal and one of the corresponding off-diagonal entries are
simultaneously different
from zero, then, also the other entries needed to span a box in the matrix
should be as well
different from zero. With this, it is easy to check that the textures
that can be generated if the $U(1)_X$ symmetry remains unbroken are not
realistic and that the best approximation
to reality comes from the two cases of a rank-one matrix, where either
all entries are allowed and equal, or only the $(3,3)$ entry is allowed.
They correspond to the choice of equal charges for all three generations:
$\alpha_1 = \alpha_2 = \alpha_3 = - \chi/2$, or to the less restrictive
choice: $\alpha_3 = - \chi/2$.
The existence of such approximate symmetries,
that are favoured by the observed hierarchy of masses in the
different fermion sectors, has been advocated
some time ago \cite{Plankl}.
Here, we will assume that, as a consequence of such a $U(1)_X$ axial
-and therefore anomalous- symmetry, above some scale $M_X$ that can
{\it a priori}
lie somewhere between the
Planck scale and an intermediate scale of the order of $10^{11}$ GeV,
only the third-generation SM fermions may acquire a mass.
This fixes the charges of $h_1$ and $h_2$ to
$\chi = - 2\alpha$, where $\alpha \equiv \alpha_3$ is the charge of the $t$
and $b$ quarks.

The next step is to investigate the possibility of generating any of the
textures, that have been proposed in ref.\cite{RRR} as candidates
for the up- and down-quark Yukawa matrices (eqs.(1,2) and Table 1),
by any mechanism that breaks this symmetry spontaneously.
The most natural scenario of this type arises in the presence of
a singlet field $\sigma$ carrying a charge $\delta$,
that couples to the two Higgs doublets through a term
$\sim \sigma h_1 h_2$ in the superpotential.
This trilinear coupling that provides a natural
explanation of the $\mu$ parameter of the MSSM,
emerges automatically in string theories \cite{Antoniadis}.
When $\sigma$ develops a vacuum expectation value (vev) $<\sigma>$,
the breaking of the symmetry can be accompanied by the
generation of higher-order Yukawa entries:
$f^c_{i} h_{1,2} ({<\sigma>\over M})^{n_{ij}} f_{j}$, where $M$ is typically
the mass of some of the Higgs particles. The power $n_{ij}$ with which
the new scale ${\cal E} = {<\sigma>\over M}$ appears in the different Yukawa
entries is such as to compensate the charge:
$x_{ij} = \alpha_i + \alpha_j - 2\alpha$ of $f^c_{i} h_{1,2} f_{j}$,
{\it i.e.}:
\begin{equation}\label{C1}
n_{ij} \delta + x_{ij} = 0 \,,
\end{equation}
where $x_{ij}$ and $\delta$ can be any positive
or negative rational numbers, while $n_{ij}$ is necessarily an integer
larger than zero.

The Yukawa matrices that can be generated in this way are proportional
to:
\begin{equation}\label{Q0}
Y_x = \left(
\begin{array}{ccc}
{\cal E}^{2 |x_1|} & {\cal E}^{|x_1 + x_2|} & {\cal E}^{|x_1|}\\
{\cal E}^{|x_1 + x_2|} & {\cal E}^{2 |x_2|} & {\cal E}^{|x_2|}\\
{\cal E}^{|x_1|} & {\cal E}^{|x_2|} & 1\\
\end{array}
\right) \,,
\end{equation}
with $x_1 = (\alpha - \alpha_1)/\delta$ and
$x_2 = (\alpha - \alpha_2)/\delta$.
In this way the connection among the different Yukawa entries becomes
transparent:
\begin{equation}
Y_{(1,1)} = Y^2 _{(1,3)}
\qquad Y_{(2,2)} = Y^2_{(2,3)}
\qquad Y_{(1,2)} = Y_{(1,3)}\cdot Y_{(2,3)}
\qquad\,.
\end{equation}
While the first relation seems to support the suppression of
the $(1,1)$ entry with respect to the $(1,3)$ entry in the matrices of
eqs.(1,2), the second relation forbids the appearance of a texture zero
in the $(2,3)$ entry and not in the $(2,2)$ entry. This poses problems
for the down-quark textures of the solutions (IV) and (V)
and for the up-quark textures of the solutions (I) and (III).
Finally, due to the third relation, it is impossible to have a zero in the
entry $(1,3)$ and simultaneously suppress
$(1,2)$ with respect to $(2,2)$ or $(2,3)$. Most of the texture structures
in eqs.(1,2) are therefore impossible to obtain in this way.
The only exception is the up-quark texture of solution (V).
If namely ${\cal E} = \lambda$, and the particular choice
$x_2 = 2$ and $x_1 = x_2^2$ is made,
\begin{equation}\label{Y}
Y_x \sim Y_u^{(V)} \sim \left(
\begin{array}{ccc}
0 & 0 & \lambda^4 \\
0 & \lambda^4 & \lambda^2 \\
\lambda^4 & \lambda^2 & 1\\
\end{array}
\right) \,,
\end{equation}
neglecting terms of order $\lambda^6$ or higher.
In order to generate the other textures another mechanism is needed.

One can, {\it e.g.}, envisage the presence of more
trilinear couplings in the superpotential containing extra singlet fields
and (heavy) Higgs multiplets that acquire masses at $M_G$.
These are very common in superstring models \cite{Antoniadis}.
Let us assume that $H_1^{-2\beta_1}$ and $H_2^{-2\beta_2}$ are two
such Higgs fields, carrying charges $-2\beta_1$ and $-2\beta_2$,
so now, the entries in the up- and down-quark Yukawa matrices will be
generated by the light Higgs fields:
\begin{equation}\label{h}
h_{u,d} \simeq h_{1,2} + \left({<\sigma_{1,2}>\over M_{1,2}}\right)^{n_{1,2}}
H_{1,2}\qquad\,,
\end{equation}
which carry a mixture of $U(1)_X$ charges.
In order to allow for a broader range of possibilities, we assume for the
moment that there are two singlet fields $\sigma_{1,2}$ with charges
$\delta_{1,2}$
that break the original symmetry subsequently
and give large masses $M_{1,2}$ to $H_1$ and $H_2$.
Defining the new scales
${\cal E}_{1,2} = {<\sigma_{1,2}>\over M_{1,2}}$,
the Yukawa textures this time become proportional
to:
\begin{equation}\label{Yz}
Y_z = \left(
\begin{array}{ccc}
{\cal E}_{1,2}^{2 |z_1|} & {\cal E}_{1,2}^{|z_1 + z_2|}
& {\cal E}_{1,2}^{|z_1 + z|}\\
{\cal E}_{1,2}^{|z_1 + z_2|} & {\cal E}_{1,2}^{2 |z_2|}
& {\cal E}_{1,2}^{|z_2 + z|}\\
{\cal E}_{1,2}^{|z_1 + z|} & {\cal E}_{1,2}^{|z_2 + z|}
& 1 + {\cal E}_{1,2}^{2 |z|}\\
\end{array}
\right) \,,
\end{equation}
whith $z_1 = (\beta_{1,2} - \alpha_1)/\delta_{1,2}$,
$z_2 = (\beta_{1,2} - \alpha_2)/\delta_{1,2}$ and $z = (\beta_{1,2} -
\alpha)/\delta_{1,2}$.

Notice that if the light and heavy Higgs bosons carry the same charges,
the power structure of $Y_z$ is the same as for $Y_x$, though the scales
may be different.
Therefore we shall assume that $z\not= 0$, in which case the relations of
eq.(8) are not valid and so the textures $Y_u^{(IV)}$
and $Y_u^{(V)}$ cannot be generated with this scenario.
Instead, now, some of the other textures of eqs.(1,2)  can be obtained.
Take first the case where $z\gg z_{1,2}$. Then the  $(1,3)$
and $(2,3)$ entries are suppressed with respect to the $(1,2)$ and $(2,2)$
entries. For the particular choice $z_2 = 2$ and $z_1 = 4$, and if
${\cal E}_1 = \lambda$ and terms of order $\lambda^8$ are neglected,
\begin{equation}
Y_z \sim Y_u^{(I)} \sim \left(
\begin{array}{ccc}
0 & \lambda^6 &  0\\
\lambda^6 & \lambda^4 & 0\\
0 & 0 & 1\\
\end{array}
\right) \,,
\end{equation}
while for the choice $z_2 = 3/2$ and $z_1 = 5/2$, and if
${\cal E}_2 = \lambda$,
\begin{equation}
Y_z \sim Y_d^{(IV),(V)} \sim \left(
\begin{array}{ccc}
0 & \lambda^4 &  0\\
\lambda^4 & \lambda^3 & 0\\
0 & 0 & 1\\
\end{array}
\right) \,.
\end{equation}
Another interesting texture structure arises when $| z_2 + z | = 2$
and $| z_1 + z_2 | = 6$, while $| z|, | z_{1,2}| \gg 0$:
\begin{equation}
Y_z \sim Y_u^{(II)} \sim \left(
\begin{array}{ccc}
0 & \lambda^6 &  0\\
\lambda^6 & 0 & \lambda^2\\
0 & \lambda^2 & 1\\
\end{array}
\right) \,.
\end{equation}

On the other hand, it is impossible to generate the textures
$Y_d^{(I)-(III)}$, because requiring $Y_{(2,2)} = Y_{(2,3)}$
would imply $z = z_2$, and therefore $Y_{(1,2)} = Y_{(1,3)}$.
However, these textures, as well as the remaining two textures
$Y_u^{(III)}$  and $Y_u^{(IV)}$, that cannot be obtained either from
$Y_x$ or from $Y_z$, can be obtained when both mechanisms are at work,
{\it i.e.} from:
\begin{equation}
Y = Y_x + Y_z \,.
\end{equation}
Let us for simplicity assume that
${\cal E} = {\cal E}_{1,2} = \lambda$.
Then, imposing the conditions: $x_2 = 2 z_2 = 3$ and $|z_1 + z_2| = 4$
when $z\gg z_2$, is sufficient to generate the first class of the missing
textures:
\begin{equation}
Y \sim Y_d^{(I)-(III)} \sim \left(
\begin{array}{ccc}
0 & \lambda^4 &  0\\
\lambda^4 & \lambda^3 & \lambda^3\\
0 & \lambda^3 & 1\\
\end{array}
\right) \,.
\end{equation}
Analogously, the choice: $x_1 = 2 z_2 = 4$ when $x_2, z_1, z\gg 0$ leads to:
\begin{equation}
Y \sim Y_u^{(III)} \sim \left(
\begin{array}{ccc}
0 & 0 & \lambda^4 \\
0 & \lambda^4 & 0 \\
\lambda^4 & 0 &  1\\
\end{array}
\right) \,,
\end{equation}
while for $x_2 = 2$ and $z_1 + z_2 = 3 x_2$ when $x_1, z_1, z_2, z\gg 0$,
we obtain:
\begin{equation}
Y \sim Y_u^{(IV)} \sim \left(
\begin{array}{ccc}
0 & \lambda^6 &  0\\
\lambda^6 & \lambda^4 & \lambda^2\\
0 & \lambda^2 & 1\\
\end{array}
\right) \,.
\end{equation}
Thus all two- and three-zero textures, that can be part of a minimal
parametrisation of the Yukawa sector at $M_G$ for the MSSM, can in
principle be generated
\footnote{It should be pointed out that the choices made for the charges
in order to generate each texture structure separately
are unique, except for cases of fine-tuning powers and scales.}
through the mixing between the two light Higgs
doublets and between two heavy Higgs fields.

Next, we look for consistent choices of the charges that can reproduce both
the up- and down-quark Yukawa textures for any of the five solutions, shown
in eqs.(1,2) and Table 1.
The first thing to notice is that if $\beta_1 = \beta_2$ and
$\delta_1 = \delta_2$, there is no such consistent choice,
since the textures in the two sectors donot differ only in scale, but
also in structure.
But even when $\beta_1 \not= \beta_2$ and/or
$\delta_1 \not= \delta_2$, it turns out that four out of the five
possible pairs of textures cannot be generated in this way, because the
conditions that need to be imposed upon the common charge variables
$x_1$ and $x_2$ are not compatible with each other, except for the solution
$(Y_u^{(II)},Y_d^{(II)})$.
For this particular case and when $\delta = \delta_1 = \delta_2$,
it is namely possible to define a set of relations
among the charges:
\begin{eqnarray}
\alpha_1 = \beta_1 - {7\over 2} \delta  & \qquad &
\alpha_2 = \alpha_1 + \delta \\
\alpha = \alpha_1 + 4 \delta & \qquad &
\beta_1 = \beta_2 + \delta
\end{eqnarray}
that lead to the textures of solution (II), eqs.(1,2).

In order to fix all the charges at least another two constraints are
needed. These can be obtained by requiring cancellation of anomalies,
or partial unification of Yukawa couplings, or both.
We start with the second option and impose the successful relations of
grand unification which relate the Yukawa couplings of the charged leptons
to those of the down quarks \cite{GJ}:
\begin{equation}
Y_{l(i,j)} = Y_{d(i,j)} \qquad  Y_{l(2,2)} = - 3 Y_{d(2,2)} \,.
\end{equation}
As a consequence, the charged leptons should carry the same $U(1)_X$
charges as the down quarks:
\begin{equation}
\gamma_1 = \alpha_1 \qquad \gamma_2 = \alpha_2 \qquad \gamma_3 = \alpha \,,
\end{equation}
though the factor that multiplies the $(2,2)$ entry of $Y_d$
cannot be accounted for by radiative mass generation,
and a different explanation is needed. Since the relations of
eqs.(19,20) are automatically satisfied, additional constraints
can come only from a neutrino sector,
containing both lefthanded and righthanded neutrino states,
$\nu_{i=1,2,3}$ and $N_{i=1,2,3}$, as this is
the case for all GUT models with the exception of $SU(5)$. These models
contain Dirac neutrino
masses that are equal to the up-quark masses \cite{Antoniadis}:
\begin{equation}
M^D_{\nu (i,j)} = M_u = Y_{u (i,j)} {<h_u>\over \sqrt{2}}
\end{equation}
and Majorana masses for the righthanded neutrinos, induced
by radiative corrections [14-16]
or by nonrenormalisable terms
\cite{nonren}:
\begin{equation}
M_{R_{i,j}} = {\cal C}\, {<H_{1,2}>^2 \over M_S}\, Y_{R(i,j)}\,,
\end{equation}
where $M_S \sim 10^{18}$ GeV is the string unification scale
and ${\cal C} \sim 1 - 10^{-3}$ is characteristic of
large-radii orbifold compactification. Since the Higgs fields acquire
a vev at $M_G$, the overall scale of the Majorana mass matrix lies
in the intermediate mass range $R \simeq 10^{11} - 10^{14}$ GeV.
The required suppression for the $\nu_{e,\mu,\tau}$ masses
is guaranteed by the familiar ``seesaw'' structure of:
\begin{equation}
M_{\nu} = \left(
\begin{array}{cc}
0 & M^D_{\nu}\\
M^{D\dag}_{\nu}& M_R \\
\end{array}
\right) \,,
\end{equation}
provided, the Majorana mass matrix $M_R$,
which for simplicity we assume to be real,
is not singular, so one can define the reduced light-neutrino mass matrix:
\begin{equation}
M_{\nu}^{eff} \simeq M_u M_R^{- 1} M_u^T  \,.
\end{equation}
The effect of the unknown Yukawa texture $Y_{R(i,j)}$  on the
light-neutrino spectra that are predicted by the five {\it Ansatz}
for the quark-mass matrices, eqs.(1,2), and the simple GUT relations,
has been studied in refs.\cite{papageor}.
The aim here is to look for consistent
choices of the fermion and scalar charges that can lead to any one of
these neutrino spectra.

Again, as a consequence of eq.(23) and the L-R symmetry, the charges for
the left- and righthanded neutrinos are the same as for the up- and down
quarks.
However, there is no need to assume that the entries in $Y_R$ are strongly
hierarchical, nor that the third-generation coupling prevails, as this is
the case in the other fermion sectors.
We can instead require all nonzero entries to be the consequence of
allowed $N^c_i H_{1,2} H_{1,2} N_j$ couplings, meaning that:
\begin{equation}
\alpha_i + \alpha_j = 2 (\beta_r + \beta_s) \,,
\end{equation}
where the indices $r,s = 1,2$ have been introduced for distinguishing
between the charges of $H_1$ and $H_2$, while $i,j$ are the generation
indices. However the eqs.(19,20) are incompatible with eq.(27) and  therefore
no masses can be generated in this way for the righthanded neutrinos.
As a matter of fact and independently of this problem, it is also
hard to reconcile the relation between the scalar
charges that follows from
eqs.(19,20): $2 \alpha = \beta_1 + \beta_2 + 2 \delta$,
with what is expected from the trilinear couplings of the
singlets to the light and heavy Higgses in the superpotential:
\begin{equation}
{\cal W} \sim \sigma^{\delta}\, h_1^{-2\alpha} h_2^{-2\alpha}\, + \,
 \sigma_{1,2}^{\delta_{1,2}}\, H_1^{-2\beta_1} H_2^{-2\beta_2}\,,
\end{equation}
namely, $2 \alpha = \beta_1 + \beta_2$, if $\delta_{1,2} = \delta \not= 0$.

Since the five quark-Yukawa texture solutions containing an asymmetric
{\it Ansatz}
cannot be generated through the ``minimal'' scenario that we have
outlined above, we would like to check the prospects for other type
of solutions where the structure of the up- and down-quark Yukawa matrices
are more alike, {\it e.g.}, they have equal number of texture zeros.
The case of three zero entries for each Yukawa matrix is ruled out
by the experimental constraints
\cite{RRR}.
The alternative case of two zeros has not been yet fully
investigated,
but the (modified) Fritzsch {\it Ansatz} still remains
the most favoured candidate:
\begin{equation}
Y_u^F \sim \left(
\begin{array}{ccc}
0 & \lambda^6 &  0\\
\lambda^6 & \lambda^4 & \lambda^2\\
0 & \lambda^2 & 1\\
\end{array}
\right) \qquad
Y_d^F \sim \left(
\begin{array}{ccc}
0 & \lambda^3 &  0\\
\lambda^3 & \lambda^2 & \lambda\\
0 & \lambda & 1\\
\end{array}
\right)
\,.
\end{equation}
Suppose now that the powers of $\lambda$ in $Y_u^F$ and in $Y_d^F$ are
generated by scales $\epsilon$ and $\epsilon^{\prime}$ that are related
through: $\epsilon^{\prime 1/2} = \epsilon = \lambda$.
Then, the two textures become identical in structure and the conditions
that lead to eq.(18) apply to each of them separately with
$\beta_1 = \beta_2$ and $\delta_1 = \delta_2$.
For the up-quark Yukawa matrix this {\it e.g.} implies the following set
of charges:
\begin{equation}
\alpha_1 \,;\quad \alpha_2 = \alpha - 2 \delta \,;\quad \alpha \,;\quad
\beta_1 = {\alpha_1 + \alpha_2 \over 2} + 3\delta_1 - \delta \,;\quad
\delta \,;\quad \delta_1 \,.
\end{equation}

If we extend this {\it Ansatz} to the neutrino sector
by means of eqs.(21,22) and require all nonzero entries to be
again the consequence of allowed $N^c_i H_{1,2} H_{1,2} N_j$ couplings,
we find the following three texture solutions for $Y_R$ that correspond
to particular
choices of the Higgs-boson charges in terms of the singlet charges
\footnote{The stars indicate nonzero entries.}:
\begin{equation}
Y^{(i)}_R \sim \left(
\begin{array}{ccc}
\star & 0 &  0\\
0 & 0 & \star\\
0 & \star & 0\\
\end{array}
\right)
\qquad
\alpha  = - 6 \delta_1 + 2 \delta
\qquad
\beta_1  = - 3 \delta_1 + {\delta \over 2}
\end{equation}
\begin{equation}
Y^{(ii)}_R \sim \left(
\begin{array}{ccc}
0 & \star &  0\\
\star & 0 & 0\\
0 & 0 & \star\\
\end{array}
\right)
\qquad \qquad \quad
\alpha = - 6 \delta_1
\qquad \quad
\beta_1  = - 3 \delta_1
\end{equation}
\begin{equation}
Y^{(iii)}_R \sim \left(
\begin{array}{ccc}
0 & 0 & \star\\
0 & \star & 0\\
\star & 0 & 0\\
\end{array}
\right)
\qquad
\alpha  = - 6 \delta_1 + 4 \delta
\qquad
\beta_1  = - 3 \delta_1 + \delta
\,.
\end{equation}
Unfortunately, again, the scalar charges donot satisfy the two conditions
implied by eq.(28). As a matter of fact, imposing the first relation
$\delta = 4\alpha$, the relative sign between
$\beta_1$ and $\delta_1$ is such that any coupling between
the singlets and the Higgs bosons in the Lagrangian will break
the $U(1)_X$ symmetry explicitly.

In view of these results, we conclude that the Yukawa textures which
have been proposed as the most economical parametrisation of the
mass spectrum of the quarks, eqs.(1,2) and Table 1,
cannot be generated in the MSSM with an
extra singlet, even, if its symmetry is extended by an extra
$U(1)$ that is broken spontaneously.
The same applies also to the case where a
pair of heavy Higgs multiplets
-coupled to an extra singlet field- are added.
It seems however that these can be generated in models with a
``richer'' scalar sector, as suggested in
refs.\cite{RRR,IR}.
We have also encountered difficulties in generating phenomenologically
acceptable Majorana mass matrices for the righthanded neutrinos
assuming a connection between the nonrenormalisable terms
needed for the latter and those needed for the generation
of the quark mass matrices.

\noindent
{\bf Acknowledgements:} I would like to thank P. Bin\'etruy
for helpful comments. This work was in part supported by the
CEC Science Projects $n^0$ SCI-CT91-0729 and CHR-XCT93-0132.

\end{document}